\documentclass[aps,prl,twocolumn,showpacs,epsfig]{revtex4-1}

\usepackage{graphicx}
\usepackage{amsmath}
\usepackage{epsfig}
\usepackage{dcolumn}
\usepackage{bm}
\usepackage{color}
\usepackage{epstopdf}

\newcommand{\ket}[1]{\left\vert#1\right\rangle}

\newcommand{\bra}[1]{\left\langle#1\right\vert}

\begin{document}
\title{Non-Markovianity of local dephasing channels and time invariant discord}

\author{P.~Haikka$^{1}$, T. H.~Johnson$^{2}$, S.~Maniscalco$^{1,3}$}

\affiliation{$^1$ Turku Center for Quantum Physics, Department of
Physics and Astronomy, University of Turku, FIN20014, Turku,
Finland} 

\affiliation{$^2$ Clarendon Laboratory, University of Oxford, Parks Road, Oxford OX1 3PU, United Kingdom} 

 \affiliation{$^3$ SUPA,
EPS/Physics, Heriot-Watt University, Edinburgh, EH14 4AS, United Kingdom}

\begin{abstract}
We study non-Markovianity and information flow for qubits experiencing local dephasing with an Ohmic class spectrum. We demonstrate the existence of a temperature-dependent critical value of the Ohmicity parameter $s$ for the onset of non-Markovianity and give a physical interpretation of this phenomenon by linking it to the form of the reservoir spectrum. We demonstrate that this link holds also for more general spectra. We unveil a class of initial states for which discord is forever frozen at a positive value. We connect time invariant discord to non-Markovianity and propose a physical system in which it could be observed.

\end{abstract}

\maketitle
{\it Introduction.-} Qubits subjected to local purely dephasing noise are ubiquitous models of open quantum systems, and they have been studied extensively in the literature. Examples include dephasing in quantum registers \cite{luczka, massimo, reina}, ultracold gases \cite{MarkusC, us}, quantum metrology protocols \cite{metrology}, quantum biological systems \cite{Qbiology}, and dynamical decoupling theory \cite{ddecoupling}. From a theoretical point of view the dephasing model is exactly solvable \cite{luczka, massimo, reina}, and hence it is an ideal testbed to investigate one of the most thrilling fields of the theory of open quantum systems, that of non-Markovian quantum processes \cite{bp}.

Recently, a great deal of attention has been devoted to the study of systems whose reduced dynamics are characterised by memory effects and recoherence phenomena, emerging from non-trivial correlations with an environment. Such dynamics are typically called non-Markovian. Memory effects and non-Markovianity have been shown to be a resource for quantum technologies \cite{metrology, QKDRuggero, NMresourse, huelgaresourse, optcontre} and consequently measures of non-Markovianity have become important as quantifiers of this resource \cite{BLP,RHP,fisher}. Moreover, it has been shown that non-Markovianity of a quantum probe can indicate a quantum phase transition occurring in a complex environment, with which the probe is interacting \cite{usLOS}.

Non-Markovian features play an important role in systems where the frequency spectrum of the environment is structured. However, a connection between the general form of the spectrum and the memory effects in the reduced system dynamics has not been elucidated until now. In this Letter we establish this connection by unveiling a necessary condition on the form of the spectrum to induce non-Markovian dynamics for a dephasing qubit. We then focus on the widely used Ohmic class of reservoir spectra and show that the condition is both necessary and sufficient for this type of spectra. Moreover, we demonstrate that only super-Ohmic environments can induce non-Markovian dynamics. This means that even if the reduced dynamics is exact, and hence no Markovian approximation has been performed, the time evolution of the qubit does not present any memory effects or recoherence for Ohmic and sub-Ohmic spectra. 

Non-Markoivan dynamics can prolong the existence of quantum properties, thus delaying the quantum to classical transition. We present here a striking example of this phenomenon by studying the decay of quantum and classical correlations in a non-Markovian open quantum system. When two qubits interact with purely dephasing local environments, the dynamics of both classical and quantum correlations can exhibit sudden changes \cite{Maziero, suddenPRL, suddenExp}; for certain initial conditions one observes a sudden transition from classical decoherence (decay of classical correlations) to quantum decoherence (decay of quantum correlations), which is preceded by a finite interval of time when quantum discord, a commonly used measure of quantum correlations \cite{OZ, HV}, is frozen to a non-zero value \cite{suddenPRL}. This scenario holds for a Markovian model of noise, that is, when one describes the system by means of a master equation of Lindblad-Gorini-Kossakowski-Sudarshan form \cite{Lindblad}, and it has also been observed for non-Markovian random telegraph noise \cite{sudden NM} and several other physical models \cite{other sudden}. 

It is natural to wonder how this transition behaves in the exact pure dephasing model considered here, and ask what the role of non-Markovianity is in this process. In this Letter we demonstrate that for this model the sudden transition does not always occur and the quantum correlations can behave in a time invariant way, that is, \emph{remaining constant at all times}, while the state of the system and all other dynamical quantities evolve. 
We present the conditions for permanently frozen discord and discuss its microscopic origin. Specifically we point out how time invariant discord is related to non-Markovian features of the exact dynamical map. 

{\it Non-Markovianity.-} Let us consider the following microscopic Hamiltonians describing the local interaction of a qubit and a bosonic reservoir, in units of $\hbar$,
 \begin{eqnarray}
 H= \omega_0 \sigma_z+ \sum_k  \omega_k a^{\dag}_k a_k + \sum_k  \sigma_z (g_k a_k+ g_k^* a^{\dag}_k), \nonumber
 \end{eqnarray}
with $\omega_0$ the qubit frequency, $\omega_k$ the frequencies of the reservoir modes, $a_k \;(a_k^{\dag})$ the annihilation (creation) operators and $g_k$ the coupling constant between each reservoir mode and the qubit. In the continuum limit $\sum_k |g_k|^2 \rightarrow \int d\omega J(\omega) \delta (\omega_k-\omega)$, where $J(\omega)$ is the reservoir spectral density. This model can be solved exactly \cite{luczka, massimo,reina}. The master equation for the qubit, in the interaction picture, is given by
\begin{eqnarray} \label{exact dissipator}
\dot{\rho}=\gamma(t) [\sigma_z \rho \sigma_z - \rho]/2, 
\end{eqnarray}
and is time-local. If the environment is initially in a thermal state, the time-dependent dephasing rate takes the form
\begin{eqnarray}
\label{gammaexact}
\gamma(t)=\int d\omega J(\omega) \coth \left[ \hbar \omega/2 k_B T\right] \sin (\omega t)/\omega,
\end{eqnarray} 
resulting in the decay of the density matrix off-diagonal elements: $\rho_{ij} (t) = e^{-\Lambda(t)} \rho_{ij} (0),\, i\neq j $, with dephasing factor $\Lambda(t) = 2 \int_0^{t} \gamma (t') dt'$ given by
\begin{eqnarray} \label{factor}
\Lambda(t) &=& 2 \int_0^{\infty} d \omega\, J(\omega)\coth \left[ \hbar \omega/2 k_B T\right] [1- \cos (\omega t)]/\omega^2\label{eq:Gamma} \nonumber\\
&\equiv& \int_0^{\infty} d \omega\, g(\omega,T) [1- \cos (\omega t)].  
\end{eqnarray}
For this model the non-Markovianity measures based on information flow \cite{BLP}, on entanglement with an ancilla \cite{RHP} and on Fisher information \cite{fisher} all predict the same crossover between Markovian and non-Markovian dynamics. The crossover is signaled by the onset of periods during which the dephasing rate is negative or, equivalently, the dephasing factor $\Lambda(t)$ always decreases with time.

To elucidate the origin and physical meaning of non-Markovianity in dephasing channels we notice that, following Ref. \cite{massimo}, one can describe the effect of the qubit on its environment by a displacement operator acting on each environment mode, with the associated phase conditional on the state of the qubit. The two qubit states excite each mode with opposing phases, leading to a decrease in the overlap between the states of the mode in each case; this is the physical cause of decoherence. Destructive interference between excitations of a mode at different times then leads to recoherences at the frequency of the mode; it is the balance between these two effects for different modes, captured exactly by Eq. (\ref{factor}), that determines whether the dynamics is non-Markovian.

Using Eq. (\ref{factor}), we establish a simple link between the onset of non-Markovianity and the form of the reservoir spectrum.  
As the cosine transform of a convex function is monotonically increasing, we deduce that a sufficient condition for Markovianity is that $g(\omega,T)$ is convex or, equivalently, the non-convexity of $g(\omega,T)$ is a necessary condition for non-Markovianity. Physically, a convex $g(\omega, T)$ means that any recoherence is always outweighed by more decoherence from lower frequency modes.
As we now show, this condition helps classify the dynamics for Ohmic-like spectral densities of the form
\begin{eqnarray}
J(\omega)=  \frac{\omega^s}{\omega_c^{s-1}} e^{-\omega/\omega_c }, \label{eq:Jomega}
\end{eqnarray}
where $\omega_c$ is the reservoir cutoff frequency.
By changing the $s$-parameter one goes from sub-Ohmic reservoirs ($s<1$) to Ohmic ($s=1$) and super-Ohmic ($s>1$) reservoirs, respectively. We stress that such  engineering of the Ohmicity of the spectrum is possible when simulating the dephasing model in trapped ultracold atoms, as demonstrated in Ref. \cite{us}. A closed analytic expression for the time-dependent dephasing rate can be found in both the zero $T$ and the high $T$  limit. In the former case one obtains
\begin{equation}\label{rate}
\gamma_0(t,s) =  [1+(\omega_c t)^2]^{-s/2} \Gamma[s] \sin \left[ s \arctan (\omega_c t)\right],
\end{equation}
with $\Gamma[x]$ the Euler gamma function. For high $T$, instead, $\gamma_{HT}(t,s) = 2 k_B T \gamma_0(t,s-1) / \omega_c $. 

Starting from Eq. (\ref{rate}) it is straightforward to prove that at zero $T$ the dephasing rate takes temporarily negative values if and only if $s>s_{crit}=2$. Hence, memory effects leading to information back flow and recoherence occur only if the reservoir spectrum is super-Ohmic with $s > 2$. Equally, Eq. (\ref{rate}) leads to $s_{crit}=3$ for high $T$.  Moreover, we have established numerically that $s_{crit}$ increases monotonically with the temperature until it reaches its maximum value $s_{crit}=3$ at infinite temperature. The existence of a temperature dependent critical value of the Ohmicity parameter, ruling the Markovian to non-Markovian transition, is one of the main results of this Letter. We now explain this result in terms of the reservoir spectrum.

It can be shown analytically that the integrand of Eq. (\ref{factor}) for the Ohmic class becomes a non-convex function of $\omega$ for $s>s_{crit}$ for both zero and infinite $T$. For intermediate temperatures, numerical investigation also indicates that the value of $s$ for which the function $g(\omega,T)$ changes from convex to non-convex coincides with $s_{crit}$, implying that the condition on the non-convexity of the spectrum is necessary and sufficient for non-Markovianity at all $T$. 
Therefore, not only does convexity guarantee decoherence always outweighs recoherence, but for these systems, it is required; this highlights the key role of the low frequency part of the spectrum in the occurrence of information backflow.

{\it Time invariant discord.-} For a bipartite state $\rho_{AB}$ the quantum discord ${\cal Q}(\rho_{AB})$ is defined as the difference between the total correlations of the system, given by the mutual information ${\cal I}(\rho_{AB})=S(\rho_A)+S(\rho_B)-S(\rho_{AB})$ with $S(\rho)=-\text{Tr}(\rho\log\rho)$ the von Neumann entropy, and the classical correlations ${\cal C}(\rho_{AB})=\max_{\Pi_A}[J(\Pi_A\rho_{AB})]$ \cite{OZ, HV}. In the latter expression the maximization is performed over all sets of orthogonal projections $\Pi_A$ on state A and $J$ describes the effect of the projective measurements on the state. In general the maximization procedure makes calculations of the quantum and classical correlations difficult, however in this work we focus on a class of states for which the maximizing measurement is known and quantum discord has an analytical expression \cite{luo}.

In Ref. \cite{suddenPRL} it was shown that if two qubits interact locally with Markovian dephasing environments, there exist classes of states for which during an initial time interval $0 < t < \bar{t}$ quantum discord remains constant while classical correlations decay. For $t > \bar{t}$, on the other hand, quantum discord decays while classical correlations remain constant. In some sense this is counter-intuitive as we would expect quantum properties to start decaying before classical properties, since they are the most sensitive to the deleterious effects of the environment. This phenomenon, named the sudden transition from classical to quantum decoherence, was investigated originally theoretically and experimentally when the qubit dynamics is described by a Markovian master equation of the form   $\dot{\rho}_{A(B)}=\gamma [\sigma^{A(B)}_j \rho_{A(B)} \sigma_j^{A(B)} - \rho_{A(B)}]/2$, with $\gamma > 0$ the constant dephasing coefficients,  $\sigma_{j}^{A(B)}$  the Pauli operator in direction $j$ acting on qubit $A (B)$, and $j=x,y,z$. For the sake of concreteness we consider dephasing along the $z$- direction only. The sudden transition occurs for initial Bell diagonal states of the form 
\begin{equation}
\rho_{AB}=\frac{(1+c)}{2}\ket{\Psi^{\pm}}\bra{\Psi^{\pm}}+\frac{(1-c)}{2}\ket{\Phi^{\pm}}\bra{\Phi^{\pm}}, \label{eq:in}
\end{equation}
where $\ket{\Psi ^{\pm}}=(\ket{00}\pm\ket{11})/\sqrt{2}$ and $\ket{\Phi^{\pm}}=(\ket{01}\pm\ket{10})/\sqrt{2}$ are the four Bell
states and $|c|<1$. The sudden transition time is given by $\bar{t}= - \ln (|c|)/(2\gamma)$. Therefore, it is possible to increase the time interval over which the discord is constant. For increasing values of  $\bar{t}$, however, the discord decreases towards its zero value obtained for $c=0$.

We now study whether the sudden transition from classical to quantum decoherence occurs also for the exact model of dephasing considered here and, if so, the effect of the reservoir spectrum on the behavior of quantum and classical correlations. For the initial class of states of Eq. (\ref{eq:in}) the mutual information and classical correlations take the form
\begin{eqnarray}\label{main}
{\cal I}[\rho_{AB}(t)]&=&\sum_{j=1}^{2}\frac{1+(-1)^{j}
c}{2}\log_{2}[1+(-1)^{j}c]\\
&+&\sum_{j=1}^{2}\frac{1+(-1)^{j}
e^{-\Lambda(t)}}{2}\log_{2}[1+(-1)^{j}e^{-\Lambda(t)}] \nonumber
\end{eqnarray}
and 
\begin{eqnarray}
{\cal C}[\rho_{AB}(t)]=\sum_{j=1}^{2}\frac{1+(-1)^{j}\chi(t)}{2}\log_{2}[1+(-1)^{j}\chi(t)], \label{cla}
\end{eqnarray}
where $\chi(t)=\max\{e^{-\Lambda(t)},c\}$, and we have taken $c$ positive for the sake of simplicity. 

From Eqs. (\ref{main})-(\ref{cla}) one sees immediately that when $e^{-\Lambda(t)} > c$ the classical correlations decay while the discord, given by the first term in Eq. (\ref{main}), remains constant. On the other hand, if a finite transition time $\bar{t}$ such that 
\begin{equation} \label{tbar}
e^{-\Lambda(\bar{t})} =  c
\end{equation}
exists, then   for $t>\bar{t}$ the discord starts decaying and the classical correlations stay constant. Contrary to the Markovian dephasing model the transition time $\bar{t}$ now crucially depends not only on $c$ but also on the parameter $s$ and on the reservoir temperature $T$ through $\Lambda(t)$.

\begin{figure}
\includegraphics[width=0.45\textwidth]{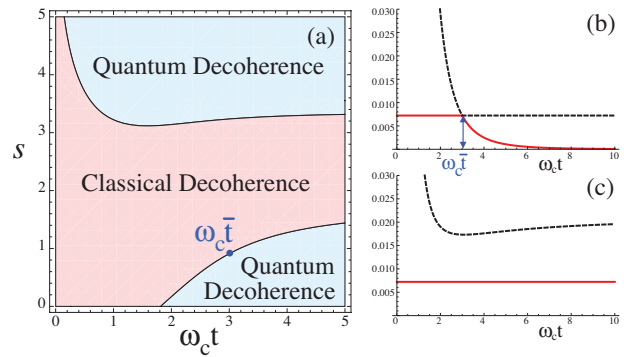}
\caption{(Color online) 
(a) Landscape of correlation dynamics in the $s-t$-plane, for $c= 0.1$ and $T=0$. Blue areas denote parameters $(t,s)$ corresponding to classical decoherence, red areas to quantum decoherence and the intersection between the two,  marking the values of $s$ and $\bar{t}$ satisfying Eq. (\ref{tbar}), defines the transition time $\bar{t}$ as a function of the reservoir spectrum parameter $s$. The two insets show discord (solid red line) and classical correlations (dashed black line) for two specific choices of $s$; (b) for $s=1$ the system has a sudden transition from quantum to classical decoherence while for (c) $s=2.5$ the discord is frozen forever. The blue dot in (a) and (b) points the transition time $\bar{t}$ for $s=1$.}
\end{figure}

Figure 1 shows the values of $s$ and $t$ for which condition (\ref{tbar}) is satisfied, for $c=0.1$ and $T=0$. For a certain range of the parameter $s$ Eq. (\ref{tbar}) has a solution and accordingly the system has a sudden transition from classical to quantum decoherence at time $\bar{t}$. Interestingly, we also discover a range of values of  $s$ for which Eq. (\ref{tbar}) has no solution and {\sl the transition time $\bar{t}$ does not exist.} For these values of $s$ only classical correlations are affected by noise, leading to classical decoherence, while discord remains frozen forever. The two different cases are illustrated in Figs. 1(b) and (c) where we plot the classical correlations and discord for the Ohmic case $s = 1$ and $s = 2.5$, respectively.

For the zero $T$ case, the possibility of having time invariant discord depends on both the initial state of the two-qubit system, that is on the parameter $c$, and on the parameter $s$, describing the structure of the reservoir spectral density. By looking at the asymptotic long time limit of condition (\ref{tbar}) we can define the $s$ and $c$ parameter space for which time invariant discord exists. This is shown in Fig. 2.  Note that the value of the frozen discord is ${\cal Q}=(1+c) \log_2(1+c)/2+(1-c)\log_2 (1-c)/2$, which is very small for small values of $c$, hence it can be argued that the phenomenon of time invariant discord is significant for larger values of $c$, roughly corresponding to $2\lesssim s\lesssim3$, for example, when the dynamics is non-Markovian. Increasing the temperature rapidly destroys the time invariant discord phenomenon. In the high temperature limit, indeed, one cannot find any value of $s$ and $c$ for which this effect occurs. We show in the following, however, that there exist realistic physical settings for which the effect can be observed. 

\begin{figure}
\includegraphics[width=0.25\textwidth]{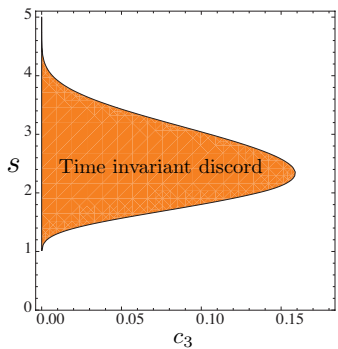}
\caption{(Color online) 
The shaded region marks range of parameters $s$ and $c$ for which the discord is frozen forever for $T=0$. Outside this region one will always observe a transition from classical to quantum decoherence.}
\end{figure}

{\it Non-Markovianity and time invariant discord.-}

Let us now link the occurrence of time invariant discord in the two qubits dynamics to the form of the reservoir spectrum and non-Markovianity in the single qubit dynamics. It is easy to convince oneself that the time invariant discord phenomenon can occur only for reservoir spectra leading to a bounded value of  $\Lambda (t)$. This ensures the existence of values of $c$ such that $e^{\Lambda(t)} > c$ for all $t$, implying that Eq.  (\ref{tbar}) is never satisfied. An asymptotic divergence of $\Lambda(t)$, on the contrary, allows for the existence of a transition time $\bar{t}$. Such a divergence, and therefore absence of time invariant discord, rests on the divergence of $g(\omega,T)/\omega$ when $\omega \rightarrow 0$  occurring for $s \le 1(2)$ at zero (finite) temperature. Similarly, convexity and thus Markovianity is ensured if $g(\omega,T)$ diverges at low frequencies, occurring for $s\le s_{crit} = 2(3)$. The above demonstrates that time invariant discord and non-Markovianity are intimately related and ultimately rely on the eventual dominance of recoherence over decoherence; thus both require the suppression of coupling to low frequency modes, embodied by the low frequency dependences of $J(\omega)$ and $g(\omega,T)$.

Finally, we note that a possible experimental setup in which the time invariant discord could be observed is an array of double-well impurities immersed in a Bose-Einstein condensate, as discussed in Refs. \cite{MarkusC,us}. When the impurities are far apart, that is, the distance $D$ between the impurities is much greater than the distance $L$ between the two potential wells forming an impurity, the collisions with the ultracold gas lead to an effective local pure dephasing model as the one considered in this Letter. For the parameters of Fig. 5 in Ref.  \cite{MarkusC}, with $D = 20 L$, the maximum value reached by the decoherence factor before attaining its stationary value is  $\max_{t} \Lambda (t) \simeq 0.058$, hence time invariant discord occurs for $0< c \lesssim 0.94$, that is for a wide range of initial states. As the discord increases for increasing values of $c$, within this physical systems one can freeze discord to values close to its maximum forever. Moreover, these systems have proven to be very resistant to the effects of finite temperature \cite{us2}, and we therefore expect to find high values of discord also for realistic temperatures of the order of $T = 10-100 {\rm nK}$.

{\it Conclusions.-} 
Non-Markovianity reflects the ability of an open system to regain and retain quantumness, since previously lost quantum information can partly flow back into the system and, in some cases, be trapped. As reservoir engineering techniques become experimentally feasible, it is crucial to establish qualitative and quantitative links between the occurrence of non-Markovianity and the form of the environmental spectrum. No such connection was known until now. In this Letter we have paved the way to these studies by presenting a necessary condition on the form of the spectrum for the non-Markovianity of a qubit undergoing pure dephasing. We have proven that, for the Ohmic-class of spectral densities the condition is necessary and sufficient, and discovered the existence of a temperature-dependent critical value of the Ohmicity parameter for the Markovian to non-Markovian crossover. For two qubits in locally dephasing environments, we have unveiled a new physical phenomenon, time invariant discord, and explored its relation to the non-Markovianity of the individual dephasing qubits.

Our results shed light on the physical origin of decoherence in a paradigmatic open quantum systems model, and are therefore of great fundamental importance. Moreover, as the model describes the dominant source of noise in several systems used for quantum technologies, our results may lead to the implementation of more resistant quantum protocols and devices.

{\it Acknowledgements.-} We acknowledge financial support from EPSRC (EP/J016349/1), the Finnish Cultural Foundation (Science Workshop on Entanglement) and the Emil Aaltonen foundation (Non-Markovian Quantum Information). We thank Dieter Jaksch for helpful discussions.

\end{document}